\documentstyle[aps]{revtex}
\begin{document}
\title{INERTIAL EFFECTS IN NON-EQUILIBRIUM THERMODYNAMICS}
\author{J.M. Rub\'{\i} and A. P\'erez-Madrid}
\address{Departament de F\'{\i}sica Fonamental,\\
Facultat de F\'{\i}sica\\
Universitat de Barcelona,\\
Diagonal 647, 08028 Barcelona, Spain\\
}
\maketitle

\begin{abstract}
We discuss inertial effects in systems outside equilibrium within the
framework of non-equilibrium thermodynamics. By introducing a Gibbs equation
in which the entropy depends on the probability density, we are able to
describe a system of Brownian particles immersed in a heat bath in both
inertial and diffusion regimes. In the former, a relaxation equation for the
diffusion current is obtained whereas in the latter we recover Fick's law.
Our approach, which uses the elements of the theory of internal degrees of
freedom, constitutes the mesoscopic version of a previous analysis which takes into account the kinetic energy of diffusion.
\end{abstract}
\pagebreak

\section{Introduction}

The macroscopic behavior of systems outside equilibrium is well described by
non-equilibrium thermodynamics \cite{kn:Mazur}. This theory, which is based
upon the local equilibrium hypothesis, provides the phenomenological
equations for the currents appearing in the conservation laws which are
necessary to establish a complete set of differential equations accounting
for the evolution of the fields. The success of the theory is a consequence
of its simplicity, its solid statistical mechanical foundations, and of the
great variety of situations to which it may be applied.

At the mesoscopic level, for times long enough, belonging to the diffusion
regime, the dynamics of a system of particles immersed in a heat bath is
governed by the Smoluchowski equation. From this equation, which involves
the probability density for the positions of the particles, we can derive
the diffusion equation accounting for the evolution of the density field.
The diffusion equation may in turn be obtained from non-equilibrium
thermodynamics by using the Fick's law in the balance of mass. If we want to
describe the dynamics at shorter time scales, governed by the corresponding
Fokker-Planck equation, the lack of equilibration in momentum space of the
particles makes it necessary to incorporate inertial terms in the
hydrodynamic description.

Our purpose in this paper is precisely to discuss how the scope of
non-equilibiurm thermodynamics goes beyond the diffusion regime and may be
extended to shorter time scales defining the inertial regime. In this
context, it was already shown in Ref. \cite{kn:Mazur} that phenomenological
equations containing the derivatives of the diffusion currents, and
consequently giving rise to relaxation equations for such currents, can be
obtained from a proper definition of the internal energy. In this
definition, the internal energy excludes the kinetic energies of all
components with respect to the centre of mass, and is then directly related
to the molecular nature of the system as it only contains contributions due
to thermal agitation and molecular interactions. After a short time, the
system enters the diffusion regime, in which the diffusion currents obey
Fick's law. This is precisely the regime in which non-equilibrium
thermodynamics has been usually applied.

Motivated by the fact that in the inertial regime the molecular nature of
the system manifests itself, and with the aim of describing the process
keeping in mind that nature, we will present in this paper a mesoscopic
treatment of the process, complementary to the macroscopic analysis, which
is based on the formulation of non-equilibrium thermodynamics in $\underline{%
\Gamma}\equiv (\vec{r},\vec{u})$-space. In this way we will be able to
obtain differential equations for the hydrodynamic fields which will
describe the behavior of the system in both diffusion and inertial regimes.
In the diffusion regime we will recover the phenomenological equation of
non-equilibrium thermodynamics whereas in the inertial regime our formalism
will give rise to a relaxation equation. Our derivation is inspired in the
introduction of internal degrees of freedom \cite{kn:Mazur}, \cite
{kn:Prigogine} to deal with the thermodynamical description of the system
when different relaxation processes, occurring in well separated time
scales, take place and constitute a more elaborated version of this theory
which is put in its proper perspective.

The paper is organized in the following way: In section II we establish the
local conservation laws which are derived from the continuity equation in $%
\underline{\Gamma}$-space. The pertinent formulation of non-equilibrium
thermodynamics is carried out in section III, whereas in section IV we
analyze the dynamics in both diffusion and inertial regimes. Finally, in the
last section we summarize our main results and clarify the scope of
non-equilibrium thermodynamics.

\section{Conservation laws}

In the hydrodynamic regime, the evolution of systems outside equilibrium is
described by a certain number of balance equations expressing the local
conservation of the hydrodynamic fields. Since we are interested in the
mesoscopic behavior of the system, our description will be carried out in
terms of the distribution function of the particles.

To proceed, we will focus on an intentionally simple and illustrative
example as is the case of a dilute suspension of Brownian particles that
will be referred to as the Brownian 'gas', which was previously introduced
in \cite{kn:Agustin}, \cite{kn:rubi} to study diffusion and thermal
diffusion processes in the context of the theory of internal degrees of
freedom. Quite generally, we will assume that, in the absence of external
forces, the single particle distribution function $f(\vec{r},\vec{u},t)$
evolves according to the conservation law in phase space 
\begin{equation}  \label{eq:a1}
\frac{\partial f}{\partial t} = -\vec{u}\cdot\nabla f -\frac{\partial}{%
\partial \vec{u}}\cdot \vec{J}_{\vec{u}}
\end{equation}
\noindent which introduces the current $\vec{J}_{\vec{u}}$ that at this
level constitutes an unspecified quantity.

The first moments of the distribution will be identified with the conserved
fields. The density of particles $\rho_B$ can then be defined as 
\begin{equation}  \label{eq:a2}
\rho_B(\vec{r},t) = m \int f(\vec{u},\vec{r},t)\; d\vec{u},
\end{equation}
\noindent where $m$ is the mass of a particle. In the same way the local
momentum of the particles, $\rho_B\vec{v}_B$, is 
\begin{equation}  \label{eq:a3}
\rho_{B} \vec{v}_B(\vec{r},t) = m \int f(\vec{u},\vec{r},t) \vec{u} \;d\vec{u%
},
\end{equation}
\noindent where $\vec{v}_B$ is the velocity field of the Brownian particles.
Finally, the internal energy of the gas $u_B$ is defined as 
\begin{equation}  \label{eq:a4}
\rho_Bu_B(\vec{r} ,t) = \frac{m}{2} \int f (\vec{u} - \vec{v}_B)^2 d \vec{u}%
\;\; .
\end{equation}

The balance equations for the hydrodynamic fields of the gas are the
evolution equations for the moments that follow by using the conservation
law (\ref{eq:a1}). Then, if we differentiate eq. (\ref{eq:a2}) with respect
to time we obtain 
\begin{equation}  \label{eq:a5}
\frac{\partial \rho_B}{\partial t} = -\nabla\cdot\rho_B\vec{v}_B
\end{equation}
\noindent where we have used eq. (\ref{eq:a3}) and the fact that $f$ goes to
zero sufficiently rapidly for large $\vec{u}$. The balance equation for the
momentum of the particles also follows from its definition (\ref{eq:a3}).
After differentiating this expression with respect to time and using again
eq. (\ref{eq:a1}) one obtains 
\begin{equation}  \label{eq:a6}
\frac{\partial}{\partial t}\rho_B\vec{v}_B = -m \int \vec{u}\vec{u} \cdot
\nabla f \; d\vec{u} + m \int \vec{J}_{\vec{u}} d \vec{u}
\end{equation}
\noindent where we have also used the asymptotic behavior of $f$.
Introducing now the pressure tensor for the particles 
\[
\vec{\vec{P}}_B = m \int f (\vec{u}-\vec{v}_B)(\vec{u}-\vec{v}_B) \; d\vec{u}
=
\]
\begin{equation}  \label{eq:a7}
m \int f \vec{u}\vec{u}\; d \vec{u} - \rho \vec{v}_B\vec{v}_B
\end{equation}
\noindent eq. (\ref{eq:a6}) transforms into 
\begin{equation}  \label{eq:a8}
\rho_B \frac{d \vec{v}_{B}}{d t} = - \nabla\cdot \vec{\vec{P}}_B(\vec{r},t)
+ m\int \vec{J}_{\vec{u}} \; d\vec{u}
\end{equation}
\noindent where we have defined the total derivative as 
\begin{equation}  \label{eq:a9}
\frac{d}{d t} \equiv \frac{\partial}{\partial t} + \vec{v}_B \cdot
\nabla\;\; .
\end{equation}
\noindent Finally, following the procedure previously indicated we can
derive the internal energy balance equation from its definition (\ref{eq:a4}%
). One obtains

\begin{equation}  \label{eq:a10}
\rho_B \frac{d u_B}{d t} = -\nabla\cdot\vec{J}_q-\vec{\vec{P}}_B : \nabla 
\vec{v}_B + m \int \vec{J}_{\vec{u}} \cdot (\vec{u}-\vec{v}_B) \;d\vec{u}%
\;\; ,
\end{equation}

\noindent where we have defined the heat flow

\begin{equation}  \label{eq:l1}
\vec{J}_q = \frac{1}{2}m\int\; f\, (\vec{u}-\vec{v}_B)^2(\vec{u}-\vec{v}%
_B)\; d\vec{u}\; .
\end{equation}

It is clear that the balance equations for the gas do not constitute, by
themselves, a closed set of differential equations since the current $\vec{J}%
_{\vec{u}}$ has not been specified. The integrals involving such a current,
appearing in eqs. (\ref{eq:a6}) and (\ref{eq:a10}), account for the
interchange of momentum and internal energy between the Brownian particles
and the heat bath, respectively. Its expression will be obtained from an
appropriate formulation of non-equilibrium thermodynamics in $\underline{%
\Gamma}$-space, which will be carried out in the next section.

\section{Non-equilibrium thermodynamics}

In the diffusion regime, non-equilibrium thermodynamics establishes the
Gibbs equation in which changes in the entropy are related to variations in
the density of the particles. Since our purpose is to carry out a
thermodynamical description of the system also in the inertial regime, in
which the molecular individuality manifests, we will generalize that idea by
assuming that variations in the probability density in $\underline{\Gamma}$%
-space, are responsible for changes in the total entropy of the system, $S$.
Our starting point will then be to formulate the Gibbs equation 
\begin{equation}  \label{eq:b1}
\delta S = - \frac{m}{T}\int \mu(\underline{\Gamma},t) \delta f(\underline{%
\Gamma},t) \;d\underline{\Gamma}
\end{equation}
\noindent where $\mu(\underline{\Gamma} ,t)$ is a chemical potential defined
in $\underline{\Gamma}$-space and $T$ the temperature of the heat bath,
assumed to be constant.

The expression of the chemical potential can be identified by requiring that
the Gibbs equation (\ref{eq:b1}) must be compatible with the Gibbs entropy
postulate \cite{kn:rubi}. This postulate establishes that 
\begin{equation}  \label{eq:h1}
S = -k\int f\; \ln f/f^{leq} \;\; d\underline{\Gamma} +S^{l.eq.}
\end{equation}

\noindent where $f$ is an arbitrary distribution function and the local
equilibrium distribution $f^{l.eq.}$ is the local Maxwellian 
\begin{equation}  \label{eq:h2}
f^{l. eq.}(\underline{\Gamma},t) = exp\left\{ \left(\mu_B(\vec{r} ,t) - 
\frac{1}{2}u^2\right)m/kT\right\}\;\; .
\end{equation}
\noindent with $\mu_B(\vec{r},t)$ being the chemical potential of the
particles at local equilibrium. Moreover, $S^{l.e.}$ is the total entropy at
local equilibrium whose variations are given by the Gibbs equation 
\begin{equation}  \label{eq:h3}
\delta S^{l.e.} = -\frac{1}{T}\int \mu_{B}(\vec{r} ,t)\delta \rho_B (\vec{r}
,t) d\vec{r}
\end{equation}

\noindent By differentiating eq. (\ref{eq:h1}) and comparing with eq. (\ref
{eq:b1}) we obtain, after using eq. (\ref{eq:h3}), the expression of the
chemical potential $\mu(\vec{u},\vec{r},t)$ that is written 
\begin{equation}  \label{eq:b5}
\mu(\vec{u},\vec{r},t) = \mu_B + \frac{kT}{m} ln\left(f/f^{l. eq.}\right)
\end{equation}

\noindent where $k$ is the Boltzmann constant. The Gibbs equation then
resembles its corresponding expression for a mixture in which the different
species would correspond to the different values of $\underline{\Gamma}$,
which may then be interpreted as an internal coordinate or degree of freedom 
\cite{kn:Mazur}.

Taking the time derivative in (\ref{eq:b1}) and using (\ref{eq:a1}) and (\ref
{eq:b5}) we obtain the entropy balance equation 
\begin{equation}  \label{eq:l1}
\frac{d S}{d t} = -\int\vec{J}_s \cdot d\vec{s}+ \int \sigma d\vec{r}
\end{equation}

\noindent where the entropy flux is given by 
\begin{equation}  \label{eq:l2}
\vec{J}_s = -k \int f (\ln f -1) \vec{u} d\vec{u}
\end{equation}

\noindent and the entropy production reads 
\begin{equation}  \label{eq:b6}
\sigma = - k \int \vec{J}_{\vec{u}}\cdot \frac{\partial}{\partial \vec{u}}
ln\left(f/f^{l. eq.}\right) \; \; d\vec{u}\;\; .
\end{equation}

\noindent This quantity is subjected to the restriction $\sigma \geq 0$,
which constitutes the formulation of the second principle. The linear laws
corresponding to eq. (\ref{eq:b6}) can be formulated by assuming isotropy
and locality in $\vec{u}$-space. One has 
\begin{equation}  \label{eq:b7}
\vec{J}_{\vec{u}} = - k L_{\vec{u}\vec{u}}\frac{\partial}{\partial \vec{u}}
ln\left(f/f^{l. eq.}\right)
\end{equation}
\noindent where $L_{\vec{u}\vec{u}}$ is a phenomenological coefficient which
may in general depend on the state variables. This expression can
alternatively be written as 
\begin{equation}  \label{eq:b8}
\vec{J}_{\vec{u}} = - \beta \left(f\vec{u} + \frac{kT}{m} \frac{\partial f}{%
\partial \vec{u}}\right)
\end{equation}
\noindent where $\beta$ is a coefficient defined as 
\begin{equation}  \label{eq:b9}
\beta\equiv mL_{\vec{u}\vec{u}}/fT
\end{equation}
\noindent which in first approximation is assumed constant. Equation (\ref
{eq:b8}) can then be used in (\ref{eq:a1}) giving 
\begin{equation}  \label{eq:b10}
\frac{\partial f}{\partial t} = -\vec{u}\cdot \nabla f + \beta \frac{\partial%
}{\partial \vec{u}}\cdot \left(f\vec{u} + \frac{kT}{m}\frac{\partial f}{%
\partial \vec{u}}\right)\; ,
\end{equation}
\noindent This expression constitutes the Fokker-Planck equation accounting
for the evolution of the distribution function of the gas.

\section{Inertial and diffusion regimes}

Having obtained the expression for the current $\vec{J}_{\vec{u}}$ from
non-equilibrium thermodynamics, we will proceed first to formulate the
complete set of hydrodynamic equations for the gas of Brownian particles.
Using eq. (\ref{eq:b8}) in (\ref{eq:a8}) and performing the corresponding
integral we then obtain 
\begin{equation}  \label{eq:c1}
\rho_B \frac{d \vec{v}_{B}}{d t} + \nabla\cdot \vec{\vec{P}}_B(\vec{r},t) =
-\beta \rho_B\vec{v}_B
\end{equation}
\noindent In the internal energy balance equation (\ref{eq:a10}), the
integral involving the current $\vec{J}_{\vec{u}}$ can also be computed by
using (\ref{eq:b8}) arriving at 
\begin{equation}  \label{eq:c4}
\rho_B \frac{d u_B}{d t} = -\nabla\cdot\vec{J}_q - \vec{\vec{P}}_B:\nabla%
\vec{v}_B - 2\beta\rho_B (u_B - u_B^{eq.})
\end{equation}
\noindent where $u_B^{eq.} = \frac{3}{2} \frac{kT}{m}$.

It becomes clear from eq. (\ref{eq:c1}) that the coefficient $\beta ^{-1}$
introduces a characteristic time scale in our system. The case $t\lesssim
\beta ^{-1}$ corresponds with the inertial regime. To discuss this regime
our starting point will be eq. (\ref{eq:c1}) that by using eq. (\ref{eq:a5})
and the definition (\ref{eq:a9}) can be expressed in terms of the diffusion
current $\vec{J}_{D}\equiv \rho _{B}\vec{v}_{B}$ in the form 
\begin{equation}
\vec{J}_{D}=-\frac{\beta ^{-1}}{1+\beta ^{-1}\nabla \cdot \vec{v}_{B}}%
\left\{ \nabla \cdot \vec{\vec{P}}_B+\frac{d\vec{J}_{D}}{dt}\right\} \,.
\label{eq:d2}
\end{equation}
\noindent This expression constitutes a relaxation equation for the
diffusion current. Up to first order in $\beta ^{-1}$, it yields 
\begin{equation}
\vec{J}_{D}=-\beta ^{-1}\left\{ \nabla \cdot \vec{\vec{P}}_{B}+\frac{d\vec{J}_{D}}{%
dt}\right\} \;,  \label{eq:d1}
\end{equation}

In order to compute the pressure tensor we need to know the form of the
distribution function. We will consider the case discussed in ref. \cite{kn:Mazur} in which the initial distribution is given by 
\begin{equation}  \label{eq:d8}
f(\underline{\Gamma},0) = exp\left\{ \left(\mu_B(\vec{r} ,0) - \frac{1}{2}(\vec{u}%
-\vec{v}_B(\vec{r} ,0))^2\right)m/kT_B\right\}\;\; .
\end{equation}
\noindent The solution of the Fokker-Planck equation at later times has also
the same Gaussian form and is given by  
\begin{equation}  \label{eq:d9}
f(\underline{\Gamma},t) = exp\left\{ \left(\mu_B(\vec{r} ,t) - \frac{1}{2}(\vec{u}%
-\vec{v}_B(\vec{r} ,t))^2\right)m/kT_B\right\}\;\; ,
\end{equation}
\noindent provided the conditions $\nabla_i v_{B\, ,i}= \frac{1}{3}
\nabla\cdot\vec{v}_B \delta_{i\, j}$ and $\nabla T_B = 0$ be fulfilled.
Using eq. (\ref{eq:d9}) in the definition of the pressure tensor given
through eq. (\ref{eq:a7}) we obtain

\begin{equation}
\vec{\vec{P}}_B=p_{B}\vec{\vec{U}}=\frac{kT_{B}}{m}\rho _{B}\;\vec{\vec{U}},  \label{eq:s0}
\end{equation}

\noindent where $\vec{\vec{U}}$ is the unit tensor. Therefore, eq. (\ref
{eq:d1}) gives

\begin{equation}  \label{eq:l4}
\vec{J}_D = -\beta^{-1}\left(\frac{kT_B}{m}\nabla\rho_B + \frac{d \vec{J}_D}{%
d t}\right) \; ,
\end{equation}

\noindent Likewise, when the distribution fuunction is given by eq. (\ref
{eq:d9}) one has $\vec{J}_q = 0$, and $u_B = \frac{3}{2}\frac{kT_B}{m}$,
which enables us to obtain the relaxation equation of the temperature $T_B$
from eq.(\ref{eq:c4})

\begin{equation}  \label{eq:s1}
\frac{d\, T_B}{d\, t} = -\frac{2}{3}T_B\nabla\cdot\vec{v}_B - 2\beta (T_B -
T)\; .
\end{equation}

\noindent Moreover, eq. (\ref{eq:c1}) transforms into

\begin{equation}
\rho _{B}\frac{d\vec{v}_{B}}{dt}=-\frac{kT_{B}}{m}\nabla \rho _{B}-\beta
\rho _{B}\vec{v}_{B}\;.  \label{eq:s2}
\end{equation}
Equations (\ref{eq:a5}), (\ref{eq:s1}), and (\ref{eq:s2}), with the equation
of state (\ref{eq:s0}), results in a closed set of differential equations
describing the evolution of the hydrodynamic fields, $\rho _{B}$, $\vec{v}%
_{B}$, $T_{B}$, and $p_{B}$.

\noindent If in addition, we assume that $T_B = T$, eq. (\ref{eq:l4}) becomes

\begin{equation}  \label{eq:l5}
\vec{J}_D = - D\nabla\rho_B -\beta^{-1}\frac{d \vec{J}_D}{d t}\; ,
\end{equation}

\noindent with

\begin{equation}  \label{eq:l10}
D = \frac{kT}{m\beta}
\end{equation}

\noindent being the diffusion coefficient. 

Inserting the relaxation equation (\ref{eq:l5}) in the balance
of mass (\ref{eq:a5}) one then obtains

\[
\frac{\partial \rho_B}{\partial t} = D \nabla^2 \rho_B - D\beta^{-1}
\nabla\cdot\frac{\partial \nabla\rho_B}{\partial t} 
\]
\begin{equation}  \label{eq:d4}
= D \nabla^2 \rho_B - D^2\beta^{-1} \nabla^2 \nabla^2\rho_B
\end{equation}
\noindent where we have neglected quadratic terms in the gradients. The
corresponding equation in $k$-space gives \cite{kn:resibois}

\begin{equation}  \label{eq:m1}
\frac{\partial \rho_B}{\partial t} = -k^2D\left(1 + D\beta^{-1}k^2\right)
\rho_B\; .
\end{equation}

\noindent These equations contains corrections of higher order in the
gradient which appear beyond the first Chapman-Enskog approximation. The
term proportional to $k^4$ in eq. (\ref{eq:m1}) corresponds to the Burnett
approximation.

When we use the distribution function (\ref{eq:d9}), with $T_B = T$, in the
expression of the entropy production given in eq. (\ref{eq:b6}), we obtain
\[
\sigma = -\frac{1}{T}\vec{J}_D\cdot\left\{ \nabla\mu_B + \frac{d}{d\, t}\vec{v}_B\right\} =
\]
\begin{equation}  \label{eq:l11}
 -\frac{1}{T}\vec{J}_D\cdot\left\{ \nabla(\mu_B + \frac{1}{2}\vec{v}%
_B^2) + \frac{\partial}{\partial\, t}\vec{v}_B\right\}\; ,
\end{equation}

\noindent where we have employed eq. (\ref{eq:a1}) and the definition of the total derivative (\ref{eq:a9}). This result agrees with
the diffusive part of the entropy production for a mixture, 

\begin{equation}\label{eq:l12}
\sigma =-\frac{1}{T}\sum_{k}\;\vec{J}_{k}\cdot \left\{ \nabla \left(\mu _{k}^{\ast}+ \frac{1}{2}\left( \vec{v}_k-\vec{v}\right)^2\right) +\frac{D%
}{Dt}\left( \vec{v}_{k}-\vec{v}\right) \right\} ,
\end{equation}

\noindent which was obtained in ref. \cite{kn:Mazur} from a standard Gibbs
equation in which the internal energy does not include the kinetic energy of diffusion of the components. Eq. (\ref{eq:l12}) corresponds with the case in which no external forces are present and $T =$ const. Here, $\vec{J}_k$ is the diffusion current of the k-th component, $\mu _{k}^{\ast}$ is the chemical potential of the k-th component,  which does not include the kinetic energy of diffusion, $\vec{v}=\rho ^{-1}\sum_{k}\rho _{k}%
\vec{v}_{k}$ is the baricentric velocity, $\rho _{k}$, and $%
\vec{v}_{k}$ are the density and velocity of the k-th component,
respectively, and $\rho $ is the total density.Moreover, the total
derivative is taken with respect to the baricentric motion
\begin{equation}\label{eq:l13}
\frac{D}{Dt}=\frac{\partial }{\partial t}+\vec{v}\cdot \nabla \, .
\end{equation}

\noindent To arrive at eq. (\ref{eq:l11}) from eq. (\ref{eq:l12}) we have to
consider that the mixture consists of two components: heat bath and Brownian
particles, with $\rho _{B}\ll \rho _{H}$ and $\vec{v}_{H}\simeq 0$,
resulting from the fact that the Brownian gas is very dilute. Here, $\rho
_{H}$ and $\vec{v}_{H}$ are the density and velocity of the heat
bath, respectively. These conditions imply that the baricentric velocity is
a negligible quantity, and consequently that $D/Dt\simeq\partial /\partial t$. Additionally, we have to identify the chemical potential of the particles in eq. (\ref{eq:l12}) with our $\mu_B$.

For $t\gg \beta ^{-1}$ one achieves the diffusion regime. Here, the inertial
term can be neglected and eq. (\ref{eq:s2}) becomes 
\begin{equation}
0=-\frac{k_{B}T}{m}\nabla \rho _{B}-\beta \rho _{B}\vec{v}_{B}  \label{eq:c5}
\end{equation}
\noindent which transforms into 
\begin{equation}
\vec{J}_{D}=-D\nabla \rho _{B}\;.  \label{eq:r1}
\end{equation}
\noindent This expression corresponds to Fick$^{^{\prime }}$s law. In this
regime, the continuity equation (\ref{eq:a5}) gives the Smoluchowski
equation 
\begin{equation}
\frac{\partial \rho _{B}}{\partial t}=D\nabla ^{2}\rho _{B}  \label{eq:c8}
\end{equation}
\noindent which describes the evolution of the density in the diffusion
regime. This equation coincides with the diffusion equation since we have
not considered interactions among particles.

To give an estimate of the characteristic time $\beta^{-1}$, let us consider
the particular situation in which the particles are spheres and the problem
is stationary. The friction coefficient per unit of mass is then given by
the Stokes law \cite{kn:Landau} 
\begin{equation}  \label{eq:d5}
\beta = \frac{6\pi\eta a}{m}\;\; .
\end{equation}

\noindent Substituting $m = (4/3)\pi a^3 \rho_p$, with $a$ being the radius
of the particle and $\rho_p$ its density, in this expression it gives 
\begin{equation}  \label{eq:d6}
\beta = \frac{9}{2} \frac{\eta}{a^2 \rho_p} = \frac{9}{2} \frac{\nu}{a^2}%
\frac{\rho_{fl}}{\rho_p}
\end{equation}
\noindent where $\rho_{fl}$ is the density of the heat bath. This equation
provides the time scale in which inertial effects should be considered. If
the fluid is water for which $\nu \sim 10^{-2} cm^3/s$ one has 
\begin{equation}  \label{eq:d7}
\beta^{-1} \sim 10^2 a^2 \frac{\rho_p}{\rho_{fl}}\;\; .
\end{equation}
\noindent Assuming that the densities are not very different and $a\sim
10^{-5} cm$, a typical particle size, one has $\beta^{-1}\sim 10^{-8} s$.
The smallness of this time then shows why non-equilibrium thermodynamics has
been so successful when analyzing transport phenomena at sufficiently long
times.

.

\section{Discussion}

In this paper we have discussed the range of applications of non-equilibrium
thermodynamics. In its more common formulation presented in \cite{kn:Mazur},
which has been successfully used to describe a wide variety of situations,
this theory implicitly assumes that linear laws relating fluxes and forces
are valid in the diffusion regime. In spite of this apparent restriction, it
is worth emphasizing that the behavior of the system in the inertial regime,
which precedes the diffusion regime, can also be described within the
framework of the theory. As pointed out in the monography \cite{kn:Mazur},
when passing from an individualized description of the components, in which
each of them has its proper velocity, to a global description characterized
dynamically by the baricentric velocity, we are neglecting the kinetic
energies of all components with respect to the center of mass, referred to
as kinetic energy of diffusion. The consideration of this energy, which as
commented in Ref. \cite{kn:Mazur} {\it is perhaps more accurate than the one
corresponding to the baricentric velocity since it only contains
contributions due to thermal agitation and molecular interactions of the
particles}, has implications in the definition of the internal
energy, since one simply introduces it by subtraction from the total
energy, kinetic and potential energies. A Gibbs equation can then be
proposed in which the entropy is as usual a function of the internal energy,
obtained by subtraction from the total energy the kinetic and potential
energies of all the components, the specific volume and the mass fractions.
The resulting entropy production contains inertial terms which give rise to
relaxation equations.

In addition to the macroscopic treatment outlined previously and with the
aim of making the analysis more in depth, we have presented in this paper a
mesoscopic version of the process based upon non-equilibrium thermodynamics
of systems with internal degrees of freedom. Maintaining the essentials of
non-equilibrium thermodynamics, we have proposed a Gibbs equation where the
entropy this time instead of depending on the density it depends on the
probability density in $\underline{\Gamma}$-space. We note in this context
that when the description is carried out in terms of the density of the
particles or of the probability density for the positions of the particles
(since they respectively evolve according to the diffusion equation and the
Smoluchowski equation) we lose information at times smaller than the
characteristic time at which the system enters the diffusion regime. In this
way we have derived the relaxation equation obtained from the macroscopic
theory which enables one to analyze the inertial and diffusion regimes. In
the inertial regime, the resulting differential equation for the density of
Brownian particles contain corrections of higher order in the density
gradient proper of the Burnett approximation.

The Gibbs equation we have proposed through eq. (\ref{eq:b1} ) is perfectly
compatible with the entropy formulated through the Gibbs postulate, by an
adequate choice of the chemical potential \cite{kn:rubi}. Apart from this
equivalence presented in section III , the Gibbs postulate, along with the
assumption that the dynamics is governed by a Fokker-Planck equation, leads
to the same entropy production, and consequently to the linear laws we have
obtained. These features provide our Gibbs equation with solid foundations.

The mesoscopic approach presented here together with the macroscopic
treatments addressed in refs. \cite{kn:Mazur} and \cite{kn:miyazaki} then
raise the question about the validity of formulating entropies including
dissipative currents as independent variables. Such a possibility, which
only may occurs when there exists a clear separation between the different
time scales \cite{kn:rubi}, is equivalent to formulate a standard Gibbs
equation in which the internal energy takes into account the individual
nature of the different components. We have corroborated this fact by means
of a mesoscopic approach in terms of the single particle distribution
function.

To end this paper it is worth pointing out that although our analysis has
been presented in the simple case of a dilute gas of Brownian particles it
could systematically be applied to more complex situations, thus arriving at
a complete description of the systems, in both inertial and diffusion
regimes, by using the method of non-equilibrium thermodynamics.

\acknowledgments
We wish to thank Prof. P. Mazur for useful discussion. This work has been
supported by DGICYT of the Spanish Government under grant PB95-0881.


\begin{references}
\bibitem{kn:Mazur}  S. R. de Groot and P. Mazur, ''Non-Equilibrium
Thermodynamics'', (Dover Publishing Co., New York, 1984). See in particular
ch. IX, $\S $ 8 and X, $\S $ 6.

\bibitem{kn:Prigogine}  I. Prigogine and P. Mazur, Physica {\bf XIX}, 241
(1953).

\bibitem{kn:Agustin}  A. P\'{e}rez-Madrid, J. M. Rub\'{\i } and P. Mazur,
Physica A, {\bf 212}, 231, (1994).

\bibitem{kn:rubi}  J.M. Rub\'{\i } and P. Mazur, Physica A {\bf 250}, (1998)
253.

\bibitem{kn:Landau}  L.D. Landau, and E.M. Lifshitz, ''Fluid Mechanics''
(2nd edn), (Pergamon, Oxford, 1987).

\bibitem{kn:resibois}  P.M.V. R\'{e}sibois, M. de Leener. {\it Classical
Kinetic Theory of Fluids}. (Wiley, New York, 1977). Ch. XII, sec. 6.3.

\bibitem{kn:miyazaki}  K. Miyazaki, K. Kitahara and D. Bedeaux, Physica A 
{\bf 230}, (1996) 600.
\end{references}
\end{document}